\pdfoutput=1

\documentclass[sigconf]{acmart}

\copyrightyear{2025}
\acmYear{2025}
\setcopyright{acmlicensed}\acmConference[ICMR '25]{Proceedings of the 2025 International Conference on Multimedia Retrieval}{June 30-July 3, 2025}{Chicago, IL, USA}
\acmBooktitle{Proceedings of the 2025 International Conference on Multimedia Retrieval (ICMR '25), June 30-July 3, 2025, Chicago, IL, USA}
\acmDOI{10.1145/3731715.3733414}
\acmISBN{979-8-4007-1877-9/2025/06}
\AtBeginDocument{%
  }

\usepackage{multirow}
\usepackage{graphicx}
\usepackage{xcolor}
\usepackage{balance}

\acmSubmissionID{437}

\begin{document}

\title{PIG: Physically-based Multi-Material Interaction with 3D Gaussians}

\author{Zeyu Xiao}
\authornote{Both authors contributed equally to this research.}
\email{xiaozy23@m.fudan.edu.cn}
\orcid{0009-0003-7522-9301}
\affiliation{
    \institution{College of Intelligent Robotics and Advanced Manufacturing, Fudan University}
    \city{Shanghai}
    \country{China}
}

\author{Zhenyi Wu}
\authornotemark[1]
\email{wuzy23@m.fudan.edu.cn}
\orcid{0009-0002-2981-9813}
\affiliation{
    \institution{College of Intelligent Robotics and Advanced Manufacturing, Fudan University}
    \city{Shanghai}
    \country{China}
}

\author{Mingyang Sun}
\email{mysun21@m.fudan.edu.cn}
\orcid{0009-0004-8217-6508}
\affiliation{
    \institution{College of Intelligent Robotics and Advanced Manufacturing, Fudan University}
    \city{Shanghai}
    \country{China}
}

\author{Qipeng Yan}
\email{qpyan23@m.fudan.edu.cn}
\orcid{0009-0002-5971-5505}
\affiliation{
    \institution{College of Intelligent Robotics and Advanced Manufacturing, Fudan University}
    \city{Shanghai}
    \country{China}
}

\author{Yufan Guo}
\email{yufanguo23@m.fudan.edu.cn}
\orcid{0009-0008-6549-9777}
\affiliation{
    \institution{College of Intelligent Robotics and Advanced Manufacturing, Fudan University}
    \city{Shanghai}
    \country{China}
}

\author{Zhuoer Liang}
\email{zeliang23@m.fudan.edu.cn}
\orcid{0009-0007-4435-9081}
\affiliation{
    \institution{College of Intelligent Robotics and Advanced Manufacturing, Fudan University}
    \city{Shanghai}
    \country{China}
}

\author{Lihua Zhang}
\email{lihuazhang@fudan.edu.cn}
\authornote{Corresponding Author.}
\orcid{0000-0003-0467-4347}
\affiliation{
    \institution{College of Intelligent Robotics and Advanced Manufacturing, Fudan University}
    \city{Shanghai}
    \country{China}
}

\renewcommand{\shortauthors}{Zeyu Xiao et al.}

\begin{abstract}
3D Gaussian Splatting has achieved remarkable success in reconstructing both static and dynamic 3D scenes. However, in a scene represented by 3D Gaussian primitives, interactions between objects  suffer from inaccurate 3D segmentation, imprecise deformation among different materials, and severe rendering artifacts. To address these challenges, we introduce PIG: Physically-Based Multi-Material Interaction with 3D Gaussians, a novel approach that combines 3D object segmentation with the simulation of interacting objects in high precision. Firstly, our method facilitates fast and accurate mapping from 2D pixels to 3D Gaussians, enabling precise 3D object-level segmentation. Secondly, we assign unique physical properties to correspondingly segmented objects within the scene for multi-material coupled interactions. Finally, we have successfully embedded constraint scales into deformation gradients, specifically clamping the scaling and rotation properties of the Gaussian primitives to eliminate artifacts and achieve geometric fidelity and visual consistency. Experimental results demonstrate that our method not only outperforms the state-of-the-art (SOTA) in terms of visual quality, but also opens up new directions and pipelines for the field of physically realistic scene generation.
\end{abstract}

\begin{CCSXML}
<ccs2012>
<concept>
<concept_id>10010147.10010371.10010396.10010400</concept_id>
<concept_desc>Computing methodologies~Point-based models</concept_desc>
<concept_significance>500</concept_significance>
</concept>
<concept>
<concept_id>10010147.10010178.10010224.10010245.10010254</concept_id>
<concept_desc>Computing methodologies~Reconstruction</concept_desc>
<concept_significance>500</concept_significance>
</concept>
<concept>
<concept_id>10010147.10010371.10010352.10010379</concept_id>
<concept_desc>Computing methodologies~Physical simulation</concept_desc>
<concept_significance>500</concept_significance>
</concept>
<concept>
<concept_id>10010147.10010178.10010224.10010245.10010247</concept_id>
<concept_desc>Computing methodologies~Image segmentation</concept_desc>
<concept_significance>300</concept_significance>
</concept>
</ccs2012>
\end{CCSXML}

\ccsdesc[500]{Computing methodologies~Point-based models}
\ccsdesc[500]{Computing methodologies~Reconstruction}
\ccsdesc[500]{Computing methodologies~Physical simulation}
\ccsdesc[300]{Computing methodologies~Image segmentation}

\keywords{Multimedia machine learning, 3D Gaussian Splatting, Physics Simulation, Scene Editing, 3D Scene Understanding}


\maketitle

\begin{figure*}
 \includegraphics[width=0.83\textwidth]{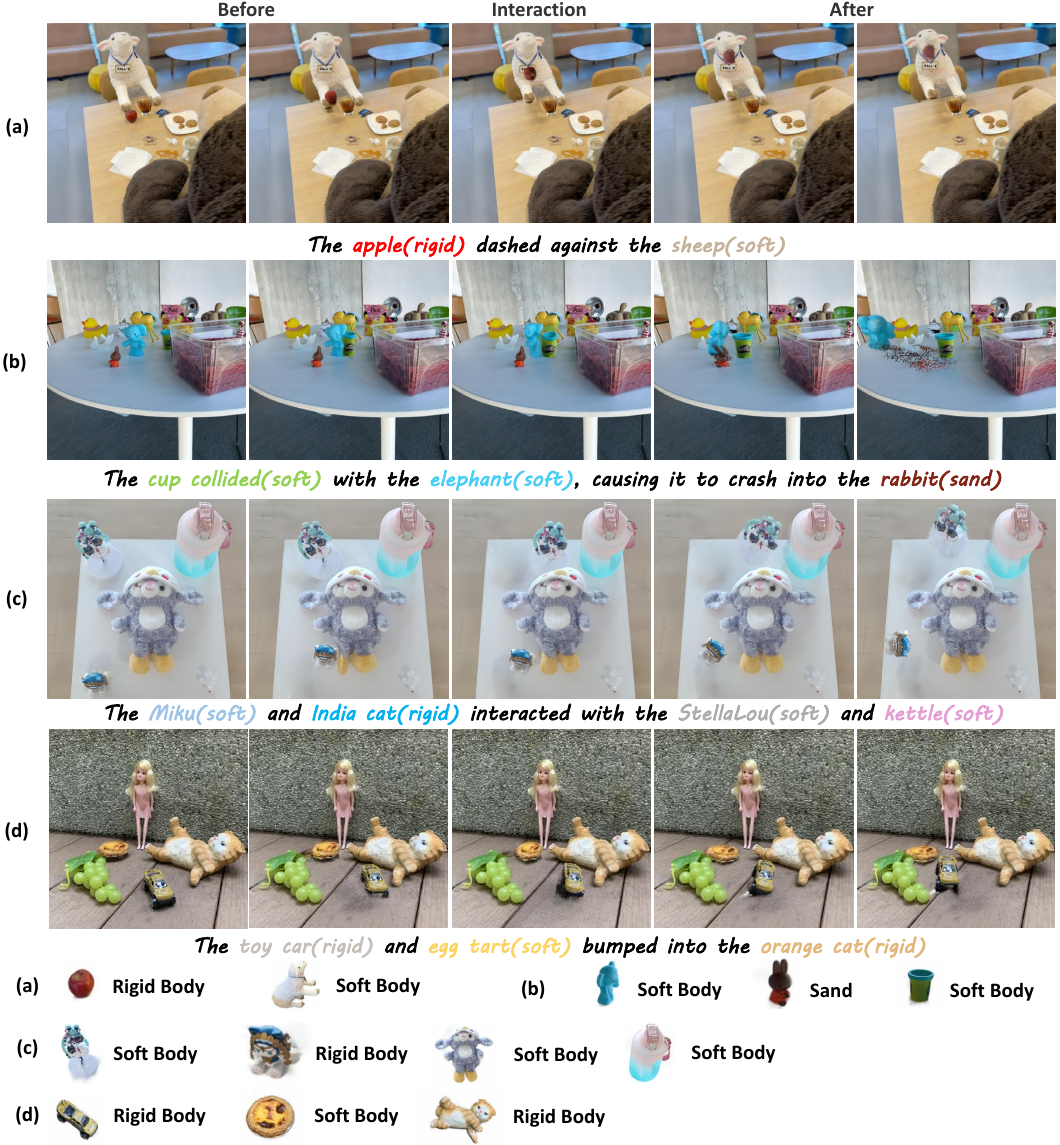}
    \centering
    \caption{We propose a novel multi-material interaction rendering pipeline, named PIG, which seamlessly integrates high-precision 3D object-level segmentation (Sec. \ref{3D Object-level Segmentation}) with physically realistic scene synthesis (Sec. \ref{Physically Realistic Scene Synthesis}). This pipeline produces physically realistic interactions between different materials across various scenes, while preserving high-quality rendering.}
    \label{all}
\end{figure*}

\section{Introduction}
\label{sec:intro}

In recent years, advancements in digital twin technology and robotic manipulation have spurred growing interest in reconstructing physically realistic scenes. Despite this progress, accurately simulating multi-material interactions within reconstructed 3D scenes while ensuring high-quality rendering remains a significant challenge. To tackle this, recent methods have integrated neural 3D scene reconstruction techniques, such as Neural Radiance Fields \cite{mildenhall2021nerf} (NeRFs) and 3D Gaussian Splatting \cite{kerbl20233d_Gaussian_splatting} (3DGS), with physical simulation approaches like the Material Point Method \cite{MPM_org} (MPM) and Mass-Spring Systems \cite{liu2013mass} (MSS). 

\begin{figure*}[t]
\centering
\includegraphics[width=0.80\textwidth]{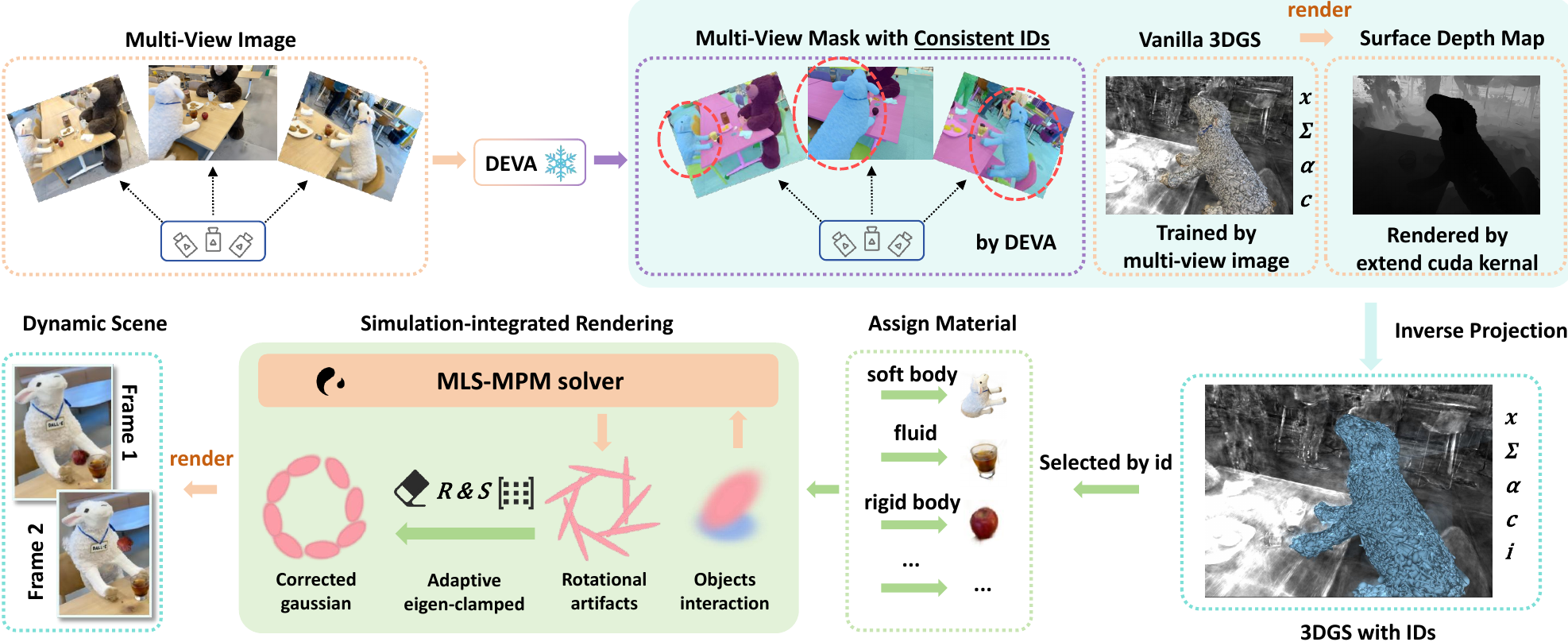} % Reduce the figure size so that it is slightly narrower than the column.
% 要求双栏的图片 slightly narrower than the column，注意调width
\caption{Pipeline of PIG. A pipeline to generate dynamic scenes for multi-material interaction from multi-view images.}
\label{pipeline}
\end{figure*}

NeRF-based approaches often rely on generating coarse meshes or embedded geometric representations \cite{peng2022cagenerf, yuan2022nerf_eding} as a prerequisite for simulation, potentially compromising accuracy. In contrast, 3DGS employs Gaussian primitives to represent 3D scenes, offering a more intuitive geometric representation and demonstrating promising results in physical simulations. For example, PhysGaussian integrates MPM with 3DGS to generate physical scenes but assumes uniform physical properties (e.g., Young's modulus and initial velocity) within user-defined regions (region level). This assumption complicates precise simulation area definition and limits applicability to multi-material interactions, particularly when multiple distinct objects exist in the chosen area. Similarly, Junhao Cai et al. \cite{massspring} incorporated a 3D mass-spring model into Gaussian primitives but assumed all scene objects to be elastic, significantly constraining material diversity in simulations. In terms of rendering, the methods mentioned above fail to address severe artifacts arising from multi-material interactions, which are significantly more critical than those occurring in scenarios limited to small deformations without interaction \cite{qiu2024feature}.

To overcome these challenges, we introduce PIG: Physically-based multi-material Interaction with 3D Gaussians. This method seamlessly integrates 3D Object-level Segmentation with Physically Realistic Scene Synthesis, effectively addressing the difficulties of multi-material interaction. To summarize, our contributions include:
\begin{itemize}
    \item We propose a novel hybrid pipeline integrating 3D object-level segmentation with high-precision particle-based simulation. By assigning distinct physical properties to objects based on segmentation results, our method operates at the object level rather than the region level, significantly enhancing simulation accuracy.
\end{itemize}
\begin{itemize}
    \item We develop an artifact removal method that applies deformation gradients to specifically correct the rotation and scaling of Gaussian primitives, making multi-material interactions appear more visually realistic. 
\end{itemize}
\begin{itemize}
    \item We conduct extensive experiments on diverse materials and datasets, demonstrating that our method achieves state-of-the-art performance in 3D object-level segmentation and artifact removal, effectively generating physically realistic multi-material interaction scenarios.
\end{itemize}

\section{Related Work}
\subsection{Segmentation with 3D Gaussian Splatting} Recent methods for segmentation using 3D Gaussian Splatting have primarily focused on distilling features from 2D pre-trained models, such as CLIP \cite{radford2021learning_CLIP}, into 3D Gaussian primitives and performs point-level segmentation. LangSplat \cite{qin2024langsplat} trained an Auto Encoder to reduce the dimensionality of features before distillation, embedding low-dimensional CLIP features into Gaussians. Segment Any 3D Gaussians \cite{cen2023saga} used 2D SAM masks \cite{kirillov2023segment} as priors and trained scene-specific features. Semantic Gaussians \cite{guo2024semantic} used Gaussian properties as inputs to train 3D features with the characteristics of 2D pre-trained models. Feature Splatting \cite{qiu2024feature} renders a feature map using DINOv2 and CLIP, trains the semantic features of each Gaussian. Our method assigns a unique ID to each object (composed of Gaussians) rather than per-Gaussian semantics, enabling complete object-level selection. Unlike point-level segmentation, which suffers from misselection and requires costly feature distillation, our approach uses multi-view mask priors with consistent IDs, eliminating extra training.

% Our method assigns an ID to each object composed of Gaussians through a customized pipeline and selects Gaussians by their ID, rather than assigning semantics to each Gaussian. In other words, object-level segmentation replaces point-level segmentation, enabling us to select an entire object completely. In contrast, point-level segmentation requires selecting each point that makes up the object sequentially through natural language, which often leads to misselections and omissions. If multiple objects in the scene are selected at the same time, in the point-level segmentation method, the same Gaussian point may belong to different objects. Additionally, the feature distillation methods mentioned above introduce extra training costs. Our method employs multi-view masks with consistent IDs as priors for segmentation without requiring additional training.

\subsection{MPM}
\label{mpm}
Material Point Method \cite{sulsky1995application} is a hybrid Lagrangian/Eulerian discretization scheme. In recent years, it has been applied to computer graphics to simulate a wide range of materials, including snow \cite{snow-MPM}, sand \cite{daviet2016semi-sand}, foam \cite{ram2015material-foam1,yue2015continuum-foam2}, cloth \cite{jiang2017anisotropic-cloth}, and solid-fluid mixtures \cite{stomakhin2014augmented-solidfluid1, tampubolon2017multi-solidfluid2}, achieving remarkable results. Since MPM uses particles to simulate physical materials, its efficiency has always been a focus of attention. Fang et al. \cite{fang2018temporally} optimized particle-grid transfers on sparse grids and explored time-step adaptivity. MLS-MPM \cite{MLS-MPM} introduced a new force computation scheme, utilizing Galerkin-style Moving Least Squares to replace the shape functions in the stress divergence term. Similarly, few works have applied the MPM for physical scene generation. For example, PhysGaussian \cite{xie2024physgaussian} proposes a method that uses MPM combined with 3DGS to generate a single-material physical scene in a selected region. This is the first method to introduce the MPM into 3DGS, but due to its lack of segmentation capability, it is challenging to accurately select the bounding boxes of simulated objects in 3D space. Furthermore, directly using the deformation gradients of particles in the MPM space to adjust the covariance of corresponding Gaussian primitives becomes ineffective when the objects undergo large deformations(Sec \ref{analysis}).
PhysDreamer \cite{zhang2024physdreamer} integrates a generative model to estimate physical parameters, which are then used for simulation through MPM. This approach heavily relies on the physical priors from the generative model and neglects the gap between simulation and rendering. Specifically, it overlooks the changes in relative positions between Gaussian primitives during the deformation process, leading to significant artifacts in the rendered objects. Unlike existing methods, we are the first to combine 3D object-level segmentation with MLS-MPM. Additionally, we eliminate artifacts through an artifacts correction method, generating physically realistic visual effects of interactions between objects of different materials. Notably, among all the approaches we have encountered so far, ours is the only one capable of utilizing 3D reconstruction techniques to assign different physical materials to the objects segmented in the reconstructed scene, thereby generating physically realistic multi-material interaction images and videos (details shown in Table \ref{table: Comparison to Concurrent Works.}).

\begin{table}[]
\caption{Comparison to Concurrent Works. It is noteworthy that our method is the only one capable of simultaneously achieving high-precision simulation and high-quality rendering within the task of multi-material interactions.}
\label{table: Comparison to Concurrent Works.}
\resizebox{0.99\linewidth}{!}{
\begin{tabular}{lccl}
\hline
Method            & \begin{tabular}[c]{@{}c@{}}Diverse \\ Physical Properties\end{tabular} & \begin{tabular}[c]{@{}c@{}}Object-Level\\ Segmentation\end{tabular} & \multicolumn{1}{c}{\begin{tabular}[c]{@{}c@{}}Artifact\\ Removal\end{tabular}} \\ \hline
PhysGaussian \cite{xie2024physgaussian}      & \multicolumn{1}{l}{}   & \multicolumn{1}{l}{}     &   \multicolumn{1}{c}{{\color{green} \checkmark}}                                                                             \\
Feature Splatting \cite{qiu2024feature} & \multicolumn{1}{l}{}     & \multicolumn{1}{l}{}            &                                                                                \\
PhysDreamer \cite{zhang2024physdreamer}       & {\color{green} \checkmark}         & \multicolumn{1}{l}{}        &                                         \\
PIG(Ours)         & {\color{green} \checkmark}        & {\color{green} \checkmark}               & \multicolumn{1}{c}{{\color{green} \checkmark}}                                 \\ \hline
\end{tabular}
}
\end{table}

\subsection{Dynamic 3D Gaussians} In recent years, many scholars have attempted to extend 3DGS to dynamic scenes. Yang et al. \cite{yang2024deformable} proposes reconstructing dynamic scenes by learning 3D Gaussians in a canonical space with a deformation field and an annealing smoothing mechanism. 4D-GS \cite{wu20244d} combines 3D Gaussians with 4D neural voxels and employs a decomposed voxel encoding alongside a lightweight MLP to predict deformations at novel timestamps, achieving real-time rendering. Gaufre \cite{liang2023gaufre} utilizes a forward-warping deformation strategy with an inductive bias-aware initialization to explicitly model non-rigid motions in dynamic regions. Dynmf \cite{kratimenos2024dynmf} decomposes per-point motion into a small set of learned basis trajectories and enforces sparsity constraints, enabling efficient real-time dynamic view synthesis. 
% These methods enable 3DGS to serve as a tool for dynamic modeling, laying the foundation for the generation of multi-material interaction scenes. However, most of these approaches fail to account for realistic physical constraints during dynamic scene reconstruction, resulting in generated scenes that often exhibit details that violate common sense.
% Compared to directly reconstructing dynamic scenes from videos, our method reconstructs a static 3D scene, segments it, and assigns physical properties, enabling more flexible and physically realistic dynamic scene generation.

These methods extend 3DGS to dynamic scenes but often lack physical constraints, leading to unrealistic results. In contrast, our approach reconstructs a static scene with physical properties, enabling more flexible and physically plausible dynamic generation.

\section{Methods}
\subsection{Preliminary.}
\subsubsection{3D Gaussian Splatting}

3DGS represents a scene as a collection of multiple Gaussian spheres suspended in 3D space, denoted as $G=\{g_i \mid i=1,2,\ldots,N\}$, where $N$ is the number of Gaussians in the scene. Each Gaussian has five trainable parameters $\{\mu_i, s_i, q_i, \alpha_i, c_i\}$, where $\mu_i \in \mathbb{R}^3$ represents the position of $g_i$ in 3D space, $s_i \in \mathbb{R}^3$ represents the scaling of $g_i$, $q_i \in \mathbb{R}^4$ is the quaternion representing the rotation of $g_i$, and $c_i$ is the color of $g_i$, which is represented in the three degrees of spherical harmonics coefficients. After representing the scene as a collection of Gaussians, the Gaussians can be depth-sorted and alpha-blended to render 2D images, which is expressed as:
\begin{equation}
\label{eq:alpha blend}
C=\sum_{i=1}^{N}T_i\alpha_ic_i,
\end{equation}
with $T_i=\prod_{j=1}^{i-1}(1-\alpha_i).$

% We further extend 3DGS by assigning an ID attribute representing an object category to each Gaussian, allowing for the selection of the target objects. Subsequently, we assign different material properties to the selected objects and perform physical simulations, achieving multi-material interaction and physically realistic scene editing.

\subsubsection{MLS-MPM} 
Traditional MPM \cite{MPM_org} discretizes the object into particles carrying various Lagrangian quantities, including position $\textbf{x}_p$, velocity $\textbf{v}_p$, and deformation gradient $F_p$. The states of the particles are then updated through the transfer operation from particles to Eulerian grids (P2G) and that from grids to particles (G2P). In this approach, during the time step from $t^n$ to $t^{n+1}$, $\nabla N_i(\textbf{x}_p)$ (27 iterations for 3D and quadratic B-spline) is used to compute the grid momentum, where $N_i(\textbf{x}_p)$ is the B-spline kernel defined on the $i$-th grid and evaluated at the particle position $\textbf{x}_p$. In our task, the real-world scenarios trained with the 3D Gaussian pipeline contains more than ten thousand Gaussian primitives, and simulating these primitives using traditional MPM methods would be time-consuming.
Therefore, we employ MLS-MPM \cite{MLS-MPM}, a particle-based physical simulation method, to simulate the segmented objects. As a hybrid Eulerian-Lagrangian approach, it updates the grid momentum using the following equation:
\begin{equation}
\frac{m_i^{n + 1}\hat {\textbf{v}}_i^{n + 1} - m_i^{n}\textbf{v}_i^{n}}{\Delta t} = m_i^{n}g + f_i^n,
\end{equation}
here, $m_i^{n}$ is the mass of the $i$-th grid, $\textbf{v}_i^{n}$ is the velocity of the $i$-th grid, $g$ is the gravity, and $\Delta t$ is the size of the time step size. Besides:
\begin{equation}
f_i^n = -\frac{\partial E}{\partial \textbf{x}_i} = - \sum \limits_p N_i(\textbf{x}_p^n) V_p^0 W_p^{-1} \frac{\partial \Psi}{\partial F}(F_p^n){F_p^n}^T (\textbf{x}_i^n - \textbf{x}_p^n) \label{2},
\end{equation}
here $i$ and $p$ represent the fields on the Eulerian grid and the Lagrangian particles respectively. $E=\sum_{p} V_{p}^{0} \Psi_{p}\left(F_{p}\right)$ where $V_p^0$ is particle initial volume and $\Psi_p$ is the energy density function, $W_p$ is the moment matrix of $\textbf{x}_p$ and $W_p = \frac{1}{4} \Delta x^2$ for quadratic $N_i(x)$ and $\frac{1}{3} \Delta x^2$ for cubic, $F_p$ is the deformation gradient of a particle from the Lagrangian view, is updated by $\mathrm{F}_{p}^{n+1}=\left(\mathrm{I}+\Delta t \mathrm{C}_{p}^{n+1}\right) \mathrm{F}_{p}^{n}$. $C_p^{n+1}$ is the affine matrix \cite{APIC} of particle p.
Compared to traditional MPM, MLS-MPM significantly reduces the computational cost while accelerating the multi-material Interaction.

\subsection{3D Object-level Segmentation.}
\label{3D Object-level Segmentation}

\subsubsection{Method Overview} 

As shown in Figure \ref{pipeline}, given a set of multi-view images $\mathcal{I}=\{I^t|t=1, 2,\dots,T\}$, where $T$ denotes the number of images in the dataset, we first reconstruct the vanilla 3D Gaussian $G=\{g_i|i=1,2,\dots,N\}$, where $N$ denotes the number of reconstructed Gaussian balls. For each $I^t \in \mathbb{R}^{3 \times H \times W}$, we employ DEVA \cite{cheng2023tracking_DEVA} to obtain a Multi-View Mask $M=\{M^t|t=1,2,\dots,T\}$ with Consistent IDs, where each element $m^t_{i,j}$ in $M^t$ represents the object ID at pixel location $(i,j)$. Meanwhile, we render the surface depth map $D=\{D^t|t=1,2,\dots,T\}$ to obtain the 2D to 3D mapping, thereby mapping the IDs from $M$ onto $G$ through inverse projection.

\subsubsection{Rendering Surface Depth Map} 
\begin{figure}[t]
\centering
\includegraphics[width=0.80\columnwidth]{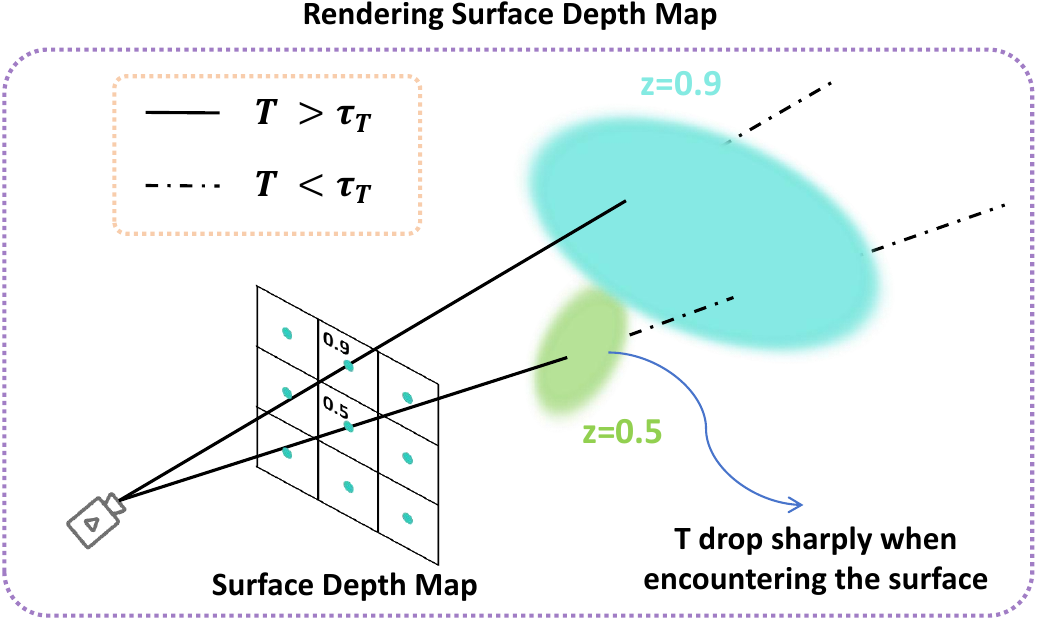} % Reduce the figure size so that it is slightly narrower than the column. Don't use precise values for figure width.This setup will avoid overfull boxes.
\caption{Illustrates of rendering the surface depth map.}
\label{rendering surface depth map}
\end{figure}

To map the $id$ from $M$ onto $G$, we identify the Gaussian corresponding to each pixel in the mask, which represents the object's surface, and obtain its depth. Then, we render it into a surface depth map. In the inverse projection, we locate the Gaussian set near this depth.
Specifically, given a mask $M^t$, for any pixel $p_{i,j}^t$, we convert it to homogeneous coordinates $P_{i,j}^t$. By using the camera's intrinsic parameters $K$ and extrinsic parameters $E$, we obtain the ray in the camera coordinate system:
\begin{equation}
    r_{i,j}^t(d)=C+dE^{-1}K^{-1}P_{i,j}^t,
\end{equation}
where $C$ is the camera's position in the world coordinate system, and $d$ represents the depth. As shown in Figure \ref{rendering surface depth map}, when $d$ increases, when the ray intersects the object surface, the transmittance $T$ in \text{Eq.} \eqref{eq:alpha blend} sharply decreases. When $T$ falls below the transmittance threshold $\tau_T$ we set, it indicates that the object surface corresponding to this pixel has been found, and we record the depth $d^t_{i,j}$ in $D^t \in \mathbb{R}^{H \times W}$.

\subsubsection{Inverse Projection} Once we obtain $D^t$, we need to find the Gaussian primitives near the corresponding depth. For each Gaussian primitive $g_i \in G$, we obtain the position of $g_i$ in the camera coordinate system $\mu_c = (x_c, y_c, z_c)$. This lets us know that $g_i$ lies on the ray $r^t_{x_c, y_c}(d)$ with a depth of $z_c$.
Next, we retrieve the depth $d^t_{x_c, y_c}$ of the pixel $p^t_{x_c, y_c}$ from $D^t$. If $z_c$ satisfies the following condition, we consider $g_i$ to correspond to the pixel $p^t_{x_c, y_c}$:
\begin{equation}
    |z_c-d^t_{x_c,y_c}|<=d^t_{x_c,y_c}\tau_d,
\end{equation}
where $\tau_d$ is a threshold that determines the range within which Gaussian primitives near the object surface are selected, we then extract the $id$ of the pixel $p^t_{x_c, y_c}$ from $M^t$ and assign it to the Gaussian selected through the mapping above process. After performing $T$ inverse projections for the $T$ camera views, each Gaussian is assumed to be mapped with $T$ ids $\{id_1, \dots, id_T\}$. We select the most frequently occurring $id$ as the final $id$ for that Gaussian.

\subsection{Physically Realistic Scene Synthesis.}
\label{Physically Realistic Scene Synthesis}

\subsubsection{Method Overview} Objects possess distinct physical properties in real-world scenarios, leading to diverse physical interactions during multi-material dynamics. In PIG, we first extract Gaussian primitives of the target objects using 3D object-level segmentation and assign material-specific physical properties. Subsequently, we utilize a simulation-integrated rendering pipeline that ensures accurate modeling of object movement and deformation while enabling artifact-free, real-time rendering, resulting in a highly physically realistic scene.

\subsubsection{Simulation-integrated rendering pipeline} Using the particle-based mechanics framework MLS-MPM, interactions can be efficiently simulated for 3D objects with corresponding physical properties segmented from 3D object-level segmentation. However, as observed in previous works (e.g., PhysGaussian \cite{xie2024physgaussian}), significant rotational artifacts degrade rendering quality. PhysGaussian modifies the particle covariance $\Sigma$ to address this issue using the deformation gradient $F_p$. Specifically, the updated covariance is computed as $\Sigma^{\prime} = F_p \Sigma F_p^T$, where $\Sigma^{\prime}$ represents the corrected covariance. Since deformation gradients primarily capture local deformation features, this method effectively handles small deformations. 

\subsubsection{Analysis of gradient-guided covariance}
\label{analysis}
In scenarios with prevalent multi-material interactions, gradient-guided covariance struggles to handle the significant deformations arising from these interactions effectively. At time $t$, $\Sigma^{\prime}_t$ can be expressed through eigenvalue decomposition into eigenvectors $Q_t$ and eigenvalues $\Lambda_t$ respectively. Using this decomposition, the gradient-guided matrices are computed as $R^\prime_t = Q_t$ and $S^\prime_t = \sqrt {\Lambda_t}$, where $R^\prime_t$ denotes the updated rotation matrix and $S^\prime_t$ the updated scaling matrix (details provided in the supplementary materials). However, under large deformation conditions, $\Sigma^\prime_t$ undergoes significant changes, causing $\sqrt {\Lambda_t}$ and $Q_t$ excessively large or small. This leads to uncontrolled deformation of Gaussian primitives, manifesting as extreme elongation, excessive compression, or even complete vanishing along certain axes, ultimately introducing severe artifacts.

\subsubsection{Adaptive eigen-clamped gaussians} To mitigate needle-like artifacts caused by excessive deformations in Gaussian primitives, we propose a straightforward yet effective method leveraging adaptive clamping based on eigenvalue decomposition. Specifically, we decompose $\Sigma^{\prime}_t$ into rotation and scaling components: $R^\prime_t = Q_t$ and $S^\prime_t = \sqrt {\Lambda_t}$. The thresholds $\tau_{\text{min}}$ and $\tau_{\text{max}}$ are then introduced to regulate scaling: if any dimension of $S^\prime_t$ for a Gaussian primitive falls below $\tau_{\text{min}}$, it is clamped to $\tau_{\text{min}}$. Similarly, if any dimension exceeds $\tau_{\text{max}}$, it is clamped to $\tau_{\text{max}}$. As a result, $S^\prime_t$ is transformed into $S_t^{\tau}$. This straightforward approach effectively resolves needle-like artifacts and significantly improves rendering quality. 

Nevertheless, due to the influence of $F_p$, all three scaling dimensions of some Gaussian primitives may be clamped to $\tau_{\text{min}}$. When rendering such primitives, they are deformed into points, potentially creating visible holes in the object. To address this issue, we define correction terms $\Delta S_t = S_t^{\tau} - S$, $\Delta R_t = R^ \prime_t - R$. The adjusted rotation and scaling matrices at time $t$ can then be expressed as:
\begin{equation}
\begin{aligned}
    R^t &= R + \lambda_R\Delta R_t, \\
    S^t &= S + \lambda_S\Delta S_t,
\end{aligned}
\end{equation}
where  $\lambda_R$ and $\lambda_S$ are constraint scales fine-tuned to optimize artifact suppression. Through extensive experimentation, we determined optimal values for $\lambda_R$ and $\lambda_S$ , achieving a balance between geometric fidelity and visual coherence. For detailed experiments on optimizing the hyperparameters, please refer to the supplementary materials.

\section{Experiments}

\begin{figure*}[t]
\centering
\includegraphics[width=0.80\textwidth]{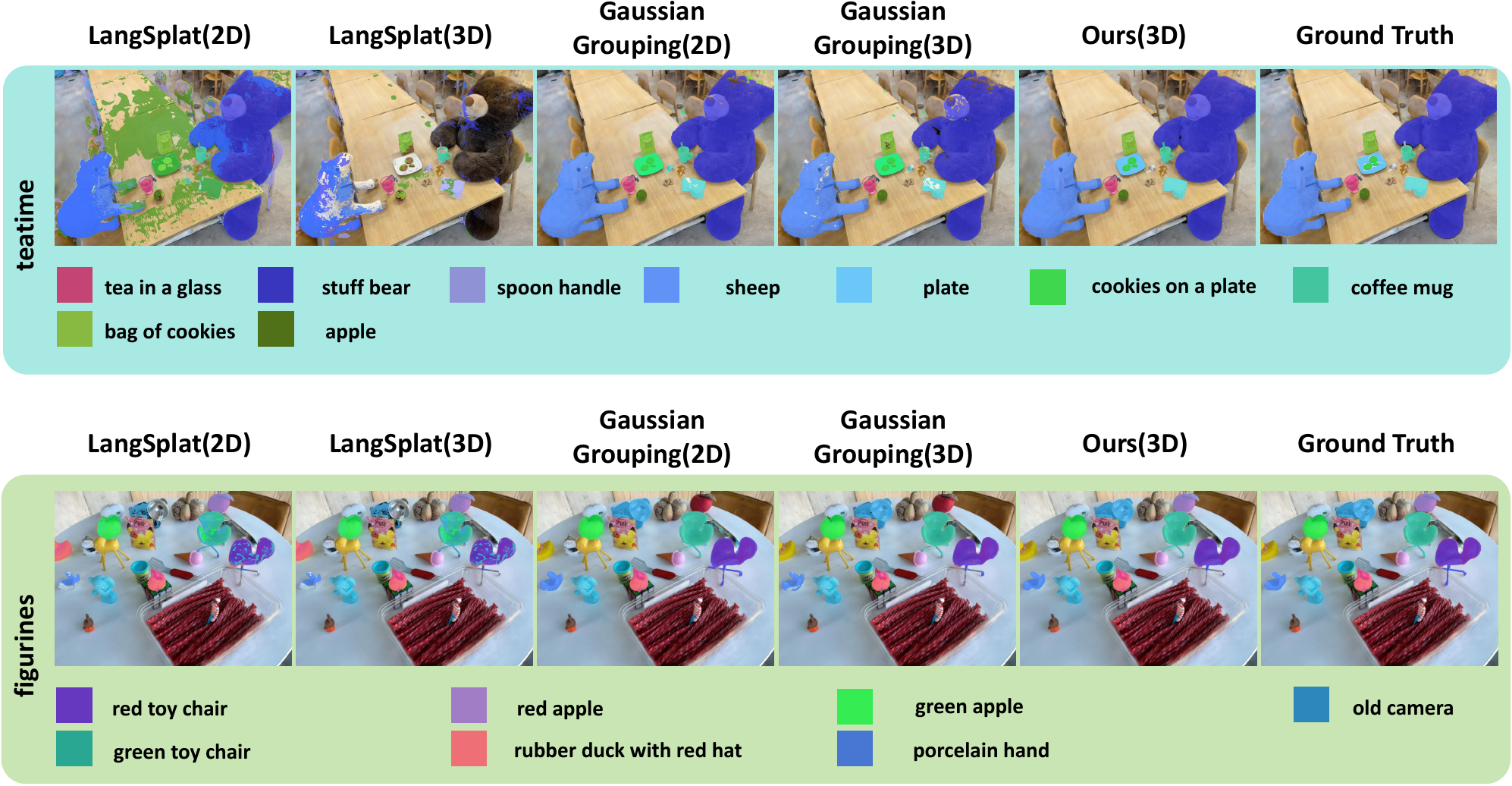} % Reduce the figure size so that it is slightly narrower than the column.
\caption{The visualization results of segmentation on the LERF-Mask dataset showcase the ``teatime'' and ``figurines'' scenes. ``2D'' represents 2D-level segmentation, while ``3D'' denotes 3D-level segmentation.}
\label{exp_segmentation_vis}
\end{figure*}

\begin{table*}[]
\caption{Quantitative comparison of segmentation quality on the 3D-OVS dataset. With ``†'' denotes 2D-level segmentation, while without ``†'' refers to 3D-level segmentation. We report the mIoU scores (\%).}
\begin{tabular}{l|ll|ll|ll|ll|ll}
\hline
\multicolumn{1}{c|}{\multirow{2}{*}{Model}} & \multicolumn{2}{c|}{bed}      & \multicolumn{2}{c|}{bench}    & \multicolumn{2}{c|}{covered\_desk} & \multicolumn{2}{c|}{blue\_sofa} & \multicolumn{2}{c}{table}     \\
\multicolumn{1}{c|}{}                       & mIoU↑         & mBIoU↑        & mIoU↑         & mBIoU↑        & mIoU↑            & mBIoU↑          & mIoU↑          & mBIoU↑         & mIoU↑         & mBIoU↑        \\ \hline
LangSplat† \cite{qin2024langsplat}                                  & 31.3          & 28.5          & \underline{94.2}          & \underline{85.3}          & \underline{92.2}             & \underline{79.7}            & 60.0           & 58.7           & 55.3          & 48.3          \\
Gaussian Grouping† \cite{ye2023gaussian_grouping}                          & \textbf{97.8} & \textbf{93.7} & 74.2          & 71.2          & 64.3             & 63.0            & \underline{83.4}           & \underline{82.9}           & \underline{89.7}          & \textbf{80.0} \\
LangSplat \cite{qin2024langsplat}                                   & 1.3           & 1.4           & 6.2           & 4.0           & 15.3             & 10.7            & 1.0            & 1.0            & 2.9           & 2.3           \\
Gaussian Grouping \cite{ye2023gaussian_grouping}                           & 96.6          & 83.6          & 73.9          & 65.6          & 63.8             & 59.4            & 82.3           & 79.8           & 89.6          & \underline{77.0}          \\
Ours                                        & \underline{97.3}          & \underline{89.2}          & \textbf{95.7} & \textbf{86.8} & \textbf{96.7}    & \textbf{85.8}   & \textbf{95.3}  & \textbf{87.4}  & \textbf{93.0} & \underline{77.0}          \\ \hline
\end{tabular}
\label{Quantitative Results of segmentation of 3D-OVS}
\end{table*}

% \usepackage{multirow}
\begin{table}[]
\caption{Quantitative comparison of segmentation quality on the LERF-Mask dataset. With ``†'' denotes 2D-level segmentation, while without ``†'' refers to 3D-level segmentation. We report the mIoU scores (\%).}
\resizebox{0.99\linewidth}{!}{
\begin{tabular}{l|ll|ll|ll}
\hline
\multicolumn{1}{c|}{\multirow{2}{*}{Model}} & \multicolumn{2}{c|}{figuriens} & \multicolumn{2}{c|}{teatime}  & \multicolumn{2}{c}{ramen}     \\
\multicolumn{1}{c|}{}                       & mIoU↑          & mBIoU↑        & mIoU↑         & mBIoU↑        & mIoU↑         & mBIoU↑        \\ \hline
LangSplat† \cite{qin2024langsplat}                                  & 55.5           & 52.6          & 50.3          & 47.9          & 38.2          & 31.1          \\
Gaussian Grouping† \cite{ye2023gaussian_grouping}                          & 69.7           & 67.9          & \underline{71.7}          & \underline{66.1}          & \textbf{77.0} & \textbf{68.7} \\
LangSplat \cite{qin2024langsplat}                                   & 11.2           & 10.4          & 27.4          & 24.9          & 13.5          & 11.2          \\
Gaussian Grouping \cite{ye2023gaussian_grouping}                           & \underline{70.8}           & \underline{68.0}          & 67.7          & 58.9          & 75.3          & 67.3          \\
Ours                                        & \textbf{91.0}  & \textbf{88.8} & \textbf{80.4} & \textbf{79.3} & \underline{75.4}          & \underline{68.0}          \\ \hline
\end{tabular}
}
\label{Quantitative Results of segmentation of LERF-MASK}
\end{table}

\begin{figure*}[t]
\centering
\includegraphics[width=0.85\textwidth]{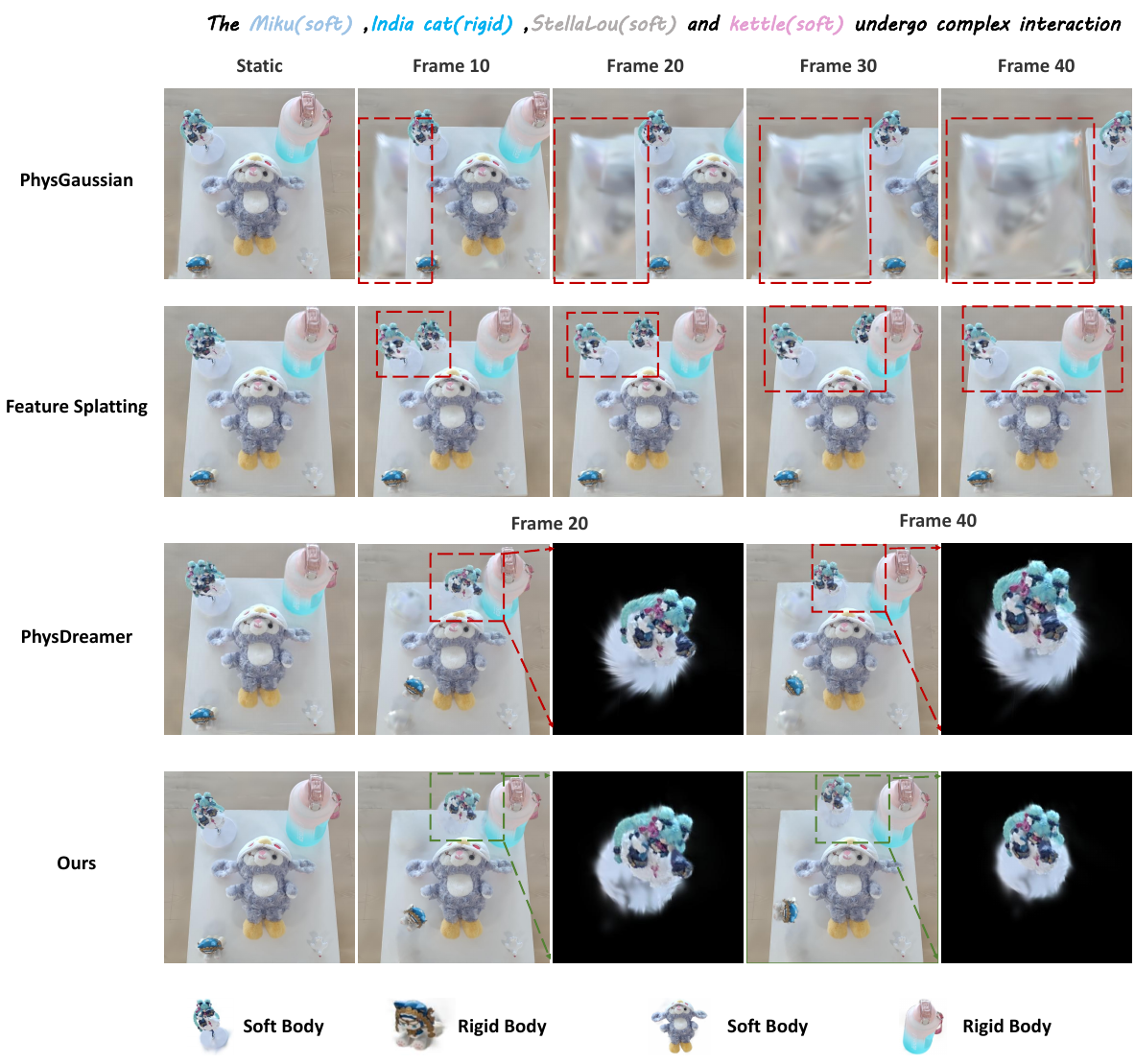} % Reduce the figure size so that it is slightly narrower than the column.
\caption{The results of a comparative experiment with several state-of-the-art methods are presented. The red rectangular areas in the figure highlight instances of physical implausibility or severe artifacts in the multi-material interaction task. Notable physical implausibility phenomena, such as object penetration (Featuresplatting \cite{qiu2024feature}) and the overall movement of the simulated region (PhysGaussian \cite{xie2024physgaussian}), are clearly visible, while artifact phenomena (PhysDreamer \cite{zhang2024physdreamer}) are relatively minor. To emphasize the capability of our method in maintaining high-fidelity rendering quality after high-precision simulation, we have zoomed in on specific regions.}
\label{compared results}
\end{figure*}

The following questions will be primarily addressed through both quantitative and qualitative experiments: 

\begin{itemize}
\item    
Sec \ref{sec: Evaluation of Segmentation in 3D Space.}: Can we segment Gaussian primitives corresponding to different objects from the scenes generated by 3DGS?
\end{itemize}
\begin{itemize}
\item   
Sec \ref{sec: Evaluation of Multi-Material Interaction}: Can we generate realistic multi-material interaction scenes from static scenes?
\end{itemize}
\begin{itemize}
\item    
Sec \ref{ablation}: Can we significantly improve rendering quality through artifact removal techniques?
\end{itemize}

% Sec \ref{sec: Evaluation of Segmentation in 3D Space.}: Can we segment Gaussian primitives corresponding to different objects from the scenes generated by 3DGS? Sec \ref{sec: Evaluation of Multi-Material Interaction}: Can we generate realistic multi-material interaction scenes from static scenes? Sec \ref{ablation}: Can we significantly improve rendering quality through artifact removal techniques?

\subsection{Evaluation of Segmentation in 3D Space.}\label{sec: Evaluation of Segmentation in 3D Space.}
\subsubsection{Task Description} 
Given a scene and the corresponding segmentation target, we generate a 2D mask. There are two methods to generate the 2D mask:
\textbf{(1)} First, use the 3DGS pipeline to render a 2D image, including an RGB image and feature map, then select the pixels corresponding to the target at the pixel level. We refer to this as 2D-level segmentation.
\textbf{(2)} Directly select the Gaussian primitives at the point level in 3D space, remove other Gaussian primitives, and render the 2D mask. We refer to this as 3D-level segmentation.
It is important to note that in our pipeline, only method \textbf{(2)} is practically meaningful, as we need to perform physical simulation directly on the Gaussian primitives.

%----------------------if edit exp--------------------
\subsubsection{Datasets and Implementation Details}
\label{dataset}
For our segmentation experiments, we employ the LERF-Mask \cite{ye2023gaussian_grouping} and 3D-OVS \cite{liu2023weakly} as the dataset. The LERF-Mask dataset includes three scenes with manually annotated ground truth masks. The 3D-OVS dataset is specifically designed for open-vocabulary 3D semantic segmentation tasks and provides a comprehensive category list. In the LERF-Mask dataset, most target objects are small-scale objects, whereas in the 3D-OVS dataset, most objects are large-scale objects. We calculate the mIoU and mBIoU between the rendered 2D masks and the manually annotated ground truth masks from the dataset. We implemented our method based on PyTorch and RTX 4070s. The hyperparameters were set as follows: $\tau_T=0.5,\tau_d=0.03$.
%----------------------if edit exp--------------------

\subsubsection{Baseline}
We compare our method with LangSplat \cite{qin2024langsplat} and Gaussian Grouping \cite{ye2023gaussian_grouping}. For 2D-level segmentation, we follow the approach outlined in the original paper. For 3D-level segmentation, we select the corresponding Gaussian using the features decoded by ``Auto Decoder'' or identity encoding in each Gaussian.

\subsubsection{Quantitative Results}
We compared the mIoU and mBIoU on the LERF-Mask and 3D-OVS dataset. As shown in Table \ref{Quantitative Results of segmentation of 3D-OVS} and Table \ref{Quantitative Results of segmentation of LERF-MASK}, our method can directly select the target Gaussians from 3D space and render 2D masks with the highest mIoU, significantly outperforming the baseline. LangSplat, which lacks any 3D priors, cannot effectively select the target Gaussians, resulting in lower 3D-level segmentation performance. In Gaussian Grouping, the presence of a 3D Regularization Loss leads to similar performance in both 2D-level and 3D-level segmentation, but it still falls short compared to our method.

\subsubsection{Qualitative Results}
To more intuitively display the segmentation quality, we create visualization results of LERF-Mask dataset as shown in Figure \ref{exp_segmentation_vis}. In 3D-level segmentation, our method provides more accurate segmentation than Gaussian grouping. For example, in the cases of ``stuff bear'' and ``sheep,'' the masks generated by Gaussian grouping contain many gaps, indicating that several Gaussians belonging to ``stuff bear'' or ``sheep'' might not have been selected. In contrast, our method selects nearly all relevant Gaussians. Even when comparing our 3D-level segmentation results to 2D-level segmentation, it is evident that our approach produces less noise and more precise boundaries.

\begin{figure}[t]
\centering
\includegraphics[width=0.95\columnwidth]{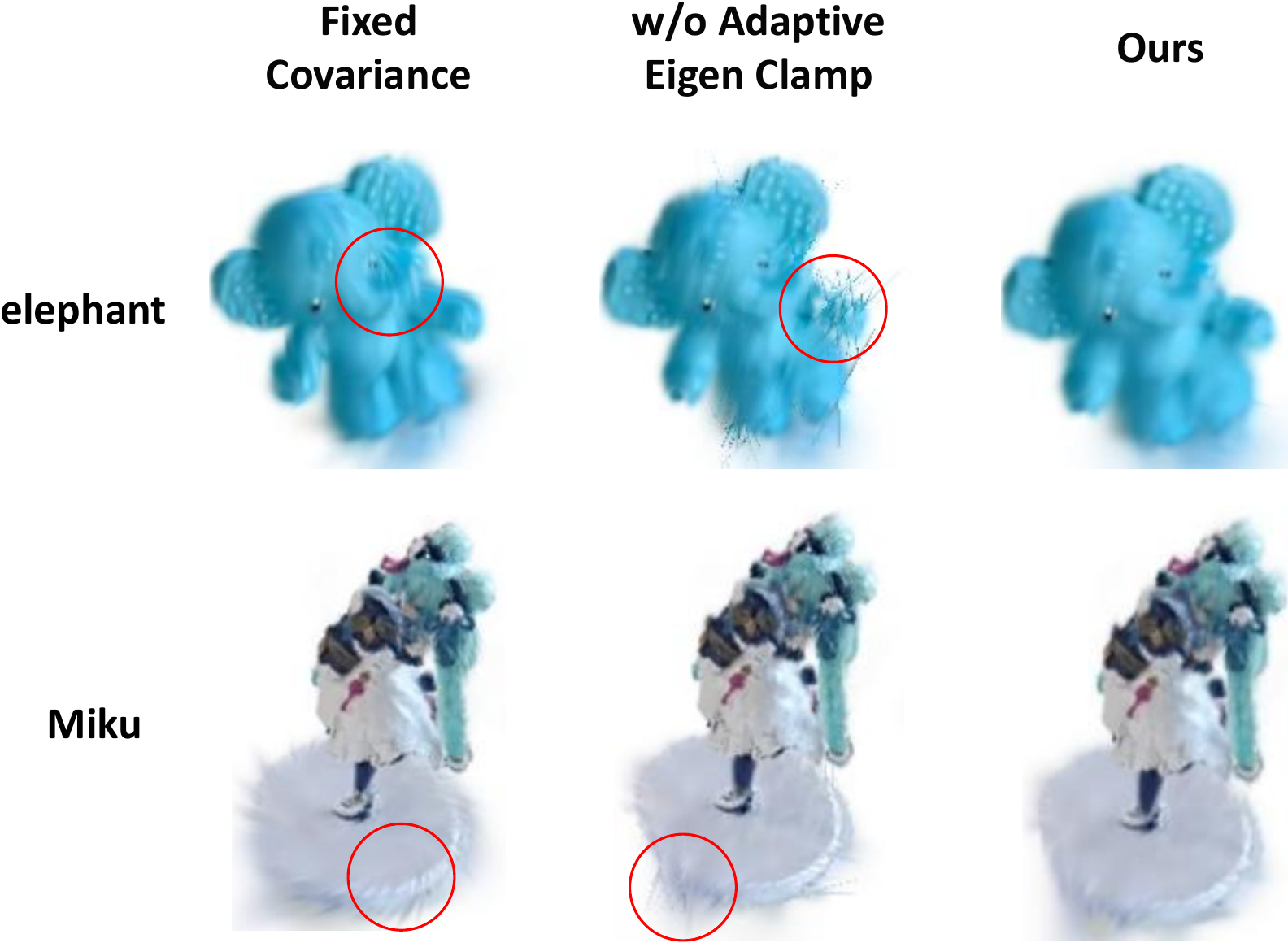} % Reduce the figure size so that it is slightly narrower than the column.
\caption{\textbf{Ablation Studies.} Inaccurate deformation of Gaussian primitives causes a large amount of artifacts. }
\label{Ablation Studies.}
\end{figure}

\subsection{Evaluation of Multi-Material Interaction}\label{sec: Evaluation of Multi-Material Interaction}

\subsubsection{Task Description.}

In a multi-material interaction task, the realism of the generated scene is primarily reflected in the following aspects: (1) Whether there is interaction between objects of two or more different materials within the scene. (2) Whether objects undergo severe motion or deformation that violates physical laws, such as penetration or self-splitting. (3) Whether artifacts are present in the video sequences, which affect the rendering quality. We designed multi-material interaction experiments in the same scene based on different baselines and rendered the corresponding visualization results to demonstrate the superiority of our pipeline in the aforementioned aspects.

\subsubsection{Setup}
\label{setup}
To evaluate the realism of multi-material interaction, in addition to the dataset used in the Sec \ref{dataset}, we also collected real-world static scenes by capturing multi-view images, which include different objects. In all subsequent experiments, we set the Young's modulus to $10^7$, the Poisson's ratio to 0.2, and the Gaussian primitives for all experimental scenes are derived from the same point cloud file. The constraint scales $\lambda_R $ and $\lambda_S$ are set to 1.2 and 0.8, respectively. The objects involved in the simulation include Miku, India Cat, StellaLou, and kettle. In the scene, the initial velocity of Miku is set to $(2, 0, 0)$, India Cat's is set to $(1, -1, 0)$, and the initial velocities of the other objects are all set to $(0, 0, 0)$.

\subsubsection{Baselines.}

We compare our pipeline with following baselines: PhysGaussian \cite{xie2024physgaussian}, PhysDreamer \cite{zhang2024physdreamer}, and Feature Splatting \cite{qiu2024feature}.
To ensure that each baseline achieves the most ideal physical property distribution, we apply specific modifications to each method. For PhysGaussian, we select the boundary positions of the Gaussian primitives involved in the simulation based on the segmentation provided by the PIG pipeline, and use these as the bounding boxes to select all Gaussian primitives within. For PhysDreamer, we initialize the physical properties of the estimated Gaussian primitives to the same values as in Sec \ref{setup}. For Feature Splatting, we use its segmentation pipeline, with the semantic label "Miku", to select the Gaussian primitives for subsequent simulation.

\subsubsection{Comparisons Results} 
We compared the multi-material interaction scenes generated by our PIG pipeline with several state-of-the-art methods. The results are presented in Fig. \ref{compared results}. Our experimental findings demonstrate that our approach significantly outperforms the current best methods in both simulation accuracy and rendering quality. Specifically, due to the reasons mentioned in Sec \ref{mpm}, in PhysGaussian, the entire simulation region moves together, which not only prevents interaction between objects of different materials but also leaves untrained regions that cannot be eliminated at their original locations. In the case of featuresplatting, only a small number of points on the object’s surface can be selected, leading to incomplete object segmentation and resulting in the formation of two separate objects: an internal part that remains stationary and an external part that suffers from severe penetration issues with the scene. PhysDreamer is the only state-of-the-art method capable of enabling object interaction in the scene, but it neglects the gap between simulation and real-world rendering. Without artifact elimination techniques, the Gaussian primitives involved in the simulation do not properly rotate or scale according to the object's deformation and displacement during rendering, leading to severe artifacts. This comparison not only highlights the substantial advantages of PIG in terms of physical realism and visual quality, but also demonstrates its tremendous potential in scene editing.

\subsection{Ablation Studies}
\label{ablation}
Given a scene and the Gaussian primitives of different objects obtained through 3D object-level segmentation, we employ a simulation-integrated rendering pipeline to simulate the deformation and displacement of objects with varying physical properties. However, this approach often introduces significant artifacts. In this section, we design extensive experiments to demonstrate the substantial improvements achieved by our method over the baseline \cite{xie2024physgaussian}.

We conducted ablation studies to intuitively highlight the necessity of Adaptive Eigen Clamp. The experimental setups include: 1) \textbf{Fixed Covariance}, where only the positions of Gaussian primitives are updated; 2) \textbf{w/o Adaptive Eigen Clamp}, which follows the approach of PhysGaussian \cite{xie2024physgaussian}, directly modifying the covariance of Gaussian primitives based on particle deformation gradients; and 3) \textbf{Ours}, which regulates Gaussian primitive deformations by adaptively clamping scaling dimensions and correcting the associated transformations.

Two examples are presented in Fig. \ref{Ablation Studies.}. The results show that without Adaptive Eigen Clamp (i.e., under Fixed Covariance or w/o Adaptive Eigen Clamp), the visual quality of the generated multi-material interactions deteriorates significantly, especially under large deformations, leading to severe rotational artifacts and needle-like distortions. In contrast, our method effectively eliminates both artifacts, achieving robust performance under both small and large deformations, surpassing the current SOTA approaches.

\section{Conclusion} This paper introduces PIG, a novel approach that successfully generates physically realistic multi-material interaction scenes, which addresses the limitations of coarse 3D segmentation, low-quality simulation, and rendering artifacts.

\textbf{Limitations.} Although our segmentation method can segment a wide range of familiar objects, it is depth-based, which often leads to incorrect segmentation for transparent and semi-transparent objects in the scene, as their accurate depth is difficult to determine.

\bibliographystyle{ACM-Reference-Format}
\balance
\bibliography{acmart}

\end{document}